\documentclass[aps,prl,reprint,groupedaddress,showpacs,amsmath]{revtex4-1}
\usepackage{mathtools}
\usepackage{graphicx}
\usepackage[colorlinks=true,linkcolor=blue,urlcolor=blue,citecolor=blue,anchorcolor=blue]{hyperref}
\usepackage{multirow}
\usepackage{pdfpages}

\newcommand{\pr}[1]{\left(#1 \right)} 
\newcommand{\br}[1]{\left[#1 \right]} 

\newcommand{\avg}[1]{\left< #1 \right>} 

\let\basetop=\top
\renewcommand{\top}{{}^\basetop \!}

\newcommand{\fa}{\: \forall \:}

\begin{document}

\title{Susceptible-infected-susceptible dynamics on the rewired configuration model}
\author{Guillaume St-Onge}
\author{Jean-Gabriel Young}
\author{Edward Laurence}
\author{Charles Murphy}
\author{Louis J. \surname{Dub\'e}}
\email[]{Louis.Dube@phy.ulaval.ca}
\affiliation{D\'epartement de Physique, de G\'enie Physique, et d'Optique, 
Universit\'e Laval, Qu\'ebec (Qu\'ebec), Canada, G1V 0A6}

\date{\today}

\begin{abstract}
We investigate the susceptible-infected-susceptible dynamics on configuration model networks. In an effort for the unification of current approaches, we consider a network whose edges are constantly being rearranged, with a tunable rewiring rate $\omega$. We perform a detailed stationary state analysis of the process, leading to a closed form expression of the absorbing-state threshold for an arbitrary rewiring rate. In both extreme regimes (annealed and quasi-static), we recover and further improve the results of current approaches, as well as providing a natural interpolation for the intermediate regimes. For any finite $\omega$, our analysis predicts a vanishing threshold when the maximal degree $k_\mathrm{max} \to \infty$, a generalization of the result obtained with quenched mean-field theory for static networks.
\end{abstract}

\pacs{64.60.aq}

\maketitle


The susceptible-infected-susceptible (SIS) dynamics is one of the classical models of disease propagation on complex networks \cite{Barrat2008,Newman2010,Pastor-Satorras2015}. It can be understood as a specific case of binary-state dynamics \cite{Gleeson2011,Gleeson2013} where nodes are either susceptible $(S)$ or infected $(I)$. Susceptible nodes become infected at rate $\lambda l$ where $l$ represents the number of infected neighbors; infected nodes recover and become susceptible at rate $\mu$, which is set to unity without loss of generality. This model has drawn much attention due to its characteristic phase transition in the stationary state ($t \to \infty)$~: It possesses an \emph{absorbing phase}---consisting of all nodes being susceptible---distinct from an \emph{active phase} where a constant fraction of the nodes remains infected on average. The former is attractive for any initial configurations with infection rate $\lambda \leq \lambda_c$, which defines the critical threshold $\lambda_c$. Despite recent advances, the theoretical understanding of this threshold for complex interaction patterns is still incomplete and remains actively pursued \cite{Pastor-Satorras2015}.

In this Letter, we examine the solution of the SIS dynamics---especially the determination of the threshold---for large size random networks with an arbitrary degree distribution $p_k$. Hence we assume, for instance, the absence of degree-degree correlation and modularity structure, which may affect the propagation on real world networks \cite{Boguna2002,Eguiluz2002}. The purpose of this contribution is to unify the approaches of the SIS dynamics on random networks under a single coherent framework and to further improve the results they provide. This is achieved by a rigorous stationary state analysis of a high-order degree-based compartmental formalism, beyond the classical mean-field approximation.

To obtain a general portrait of the situation, our perspective is to consider the structure as a dynamical object---a \emph{rewiring network}. The structure is evolving according to a continuous Markov chain, independent of the SIS dynamics~: At rate $\omega M/2$, where $M$ is the total number of edges in the network, two random edges are chosen and the stubs are rematched. This process samples the configuration model---multi-edges and self-loops are allowed---and leaves the degree sequence unaltered \cite{Fosdick2016}. The rewiring rate $\omega \geq 0$ permits us to easily tune the interplay between the disease propagation and the structural dynamics.

\begin{table}
\caption{Threshold estimates for different rewiring regimes based on current formalisms \cite{Pastor-Satorras2001,VanMieghem2009,Mata2014}.\label{tab:threshold}}
\begin{ruledtabular}
\begin{tabular}{l l l}
Regime & Formalism & Threshold estimate $\lambda_c$\\
\hline 
 \multirow{2}{*}{Quasi-static} & QMF 	  & $1/\mathrm{max}\pr{\sqrt{k_\mathrm{max}}, \avg{k^2}/\avg{k}}$ \\
 & PHMF      & $\avg{k}/\pr{\avg{k^2} - \avg{k}}$ \\
 Annealed & HMF 	  & $\avg{k}/\avg{k^2}$
\end{tabular}
\end{ruledtabular}
\end{table}

\paragraph*{Overview of existing results.} Current approaches to the SIS dynamics yield approximate solutions to this dual process for two extreme regimes (see table \ref{tab:threshold}). There is the \emph{annealed} network regime when the rewiring is much faster than the propagation dynamics ($\omega\to \infty$). The \emph{dynamical correlation}---the state correlation between pairs of nodes---is suppressed, simplifying greatly the analysis. An exact solution for the stationary state of the dynamics is obtained with the heterogeneous mean-field theory (HMF) \cite{Pastor-Satorras2001a,Pastor-Satorras2001} from which $\lambda_c$ is explicitly calculated (see table \ref{tab:threshold}).

In contrast, there is the \emph{quasi-static} network regime ($\omega \to 0$). Between each rewiring event, the SIS dynamics has enough time to relax and reach the stationary state---temporal averages for the dynamics are then equivalent to ensemble averages on every static realization of the configuration model. A typical approach to study this regime is quenched mean-field theory (QMF) \cite{Pastor-Satorras2015,VanMieghem2009}, which solves the SIS dynamics for a static network realization, taking into account explicitly the whole structure through the adjacency matrix. The threshold (table \ref{tab:threshold}) is then expressed in terms of the largest eigenvalue of the adjacency matrix $\Lambda_1 \sim \mathrm{max}\pr{\sqrt{k_\mathrm{max}}, \avg{k^2}/\avg{k}}$. Contrary to the annealed regime, dynamical correlation has a large impact on the dynamics in the quasi-static regime. To treat this aspect in details, standard QMF has been improved with moment closures for pairs of nodes (pair QMF) and leads, notably, to more accurate results for the evaluation of the threshold \cite{Cator2012,Mata2013}.

Another approach to study the quasi-static regime would be the pair heterogeneous mean-field theory (PHMF) \cite{Mata2014,Cai2016}. This degree based approach is an extension of HMF, still considering that every node is coupled to a same mean-field, but where dynamical correlation is considered through a pair approximation scheme. Even though describing the same regime, PHMF and QMF do not agree for every distribution $p_k$ in the thermodynamic limit (number of nodes $N \to \infty$) \cite{Ferreira2012}. In fact, we will provide below compelling evidences to show that PHMF is not accurate for certain degree distributions.

\paragraph*{The rewired network approach (RNA).} In degree-based approaches, the probability that a node of degree $k$ is infected, denoted $ \rho_k(t)$, follows the rate equation \cite{Pastor-Satorras2015}
\begin{align}\label{eqRhoClosed}
	\dot{\rho_k} &= - \rho_k + \lambda k (1- \rho_k) \theta_k \;,
\end{align}
where $\theta_k(t)$ is the probability of reaching an infected node following a random edge starting from a degree $k$ susceptible node. In the stationary state ($\dot{\rho}_k = 0 \; \fa k$), the following relation
\begin{align}\label{rhoSS}
	\rho_k^* = \frac{\lambda k \theta_k^* }{1 + \lambda k \theta_k^* } \quad \text{or} \quad \lambda k \theta_k^* = \frac{\rho_k^*}{1- \rho_k^*} \;,
\end{align}
is obtained. Stationary values will be marked hereafter with an asterisk (*). Since the formalism derives its accuracy from $\theta_k^*$, our objective is to find the most precise expression of this probability, taking into account the rewiring process. 

To do so, we adapt the approach proposed in Refs.~\cite{Marceau2010,Gleeson2011}, which consists of a set of differential equations governing the evolution of the compartments of nodes of a specified degree and infected degree (see also Refs.~\cite{Lindquist2011,Gleeson2013}). To incorporate the rewiring process, we introduce the probability that a newly rewired stub reaches an infected node $\Theta(t) \equiv \avg{k \rho_k}/\avg{k}$ (all averages are taken over $p_k$). Let $s_{kl}(t)$ [$i_{kl}(t)$] be the probability that a degree $k$ node is susceptible (infected) and has $l \leq k$ infected neighbors. The rate equations for these probabilities are\begin{subequations}\label{ame}
\begin{align}
	\dot{s_{kl}} =& i_{kl} - \lambda l s_{kl}  + [1+\omega(1- \Theta)]\br{(l+1)s_{k(l+1)} - l s_{kl}}   \nonumber \\ +& (\Omega^S + \omega \Theta) \br{(k-l+1)s_{k(l-1)} - (k-l)s_{kl}} \;, \label{ame1}\\
	\dot{i_{kl}} =&  \lambda l s_{kl} - i_{kl} + [1+\omega(1- \Theta)]\br{(l+1)i_{k(l+1)} - l i_{kl} } \nonumber \\ +& (\Omega^I + \omega \Theta) \br{(k-l+1)i_{k(l-1)} - (k-l)i_{kl}} \;, \label{ame2}
\end{align}
\end{subequations}
where $\Omega^S(t)$ and $\Omega^I(t)$ are the mean infection rates for the neighbors of susceptible and infected nodes. These rates can be estimated from the compartmentalization \cite{Gleeson2011}, yielding
\begin{align}\label{defOmega}
	\Omega^S &= \lambda \frac{\sum_{k,l}(k-l)ls_{kl}p_k}{\sum_{k,l}(k-l)s_{kl}p_k} \;, & \Omega^I &= \lambda \frac{\sum_{k,l}l^2s_{kl}p_k}{\sum_{k,l}l s_{kl}p_k} \;.
\end{align}

One notes that by definition $\sum_l l s_{kl} = (1- \rho_k)k \theta_k$. Hence, a rate equation for $\theta_k$ can be defined using Eqs.~\eqref{ame} (see the Supplemental Material \cite{SM}). To obtain a closed solution for $\theta_k^*$ in the stationary state limit, we use the \emph{pair approximation}
\begin{align}\label{2ndMoment}
 \sum_{l=0}^k l^2 s_{kl}^* \approx (1- \rho_k^*) \br{k \theta_k^* + k (k-1) {\theta_k^*}^2} \;,
\end{align}
which expresses that the state of each neighbor is independent, as proposed in Ref.~\cite{Gleeson2011}. Under this approximation, we find \cite{SM}
\begin{align}\label{thetaSol}
	\theta_k^*(\omega,\lambda) &= 	\begin{dcases}
					\frac{\beta}{\kappa - 1} &\text{if } k = 1 \;,\\
					\frac{k - \kappa + \sqrt{(k - \kappa)^2 + 4 \alpha \beta(k-1)}}{2 \alpha	(k-1)} &\text{if } k > 1 \;,
				\end{dcases}
\end{align}
with the parameters defined as
\begin{align}
	\alpha &= \frac{1 + \omega + {\Omega^I}^* }{{\Omega^I}^*+ \omega \Theta^*} \;, \quad \beta  = \frac{({\Omega^S}^* + \omega \Theta^*)(2 + \omega + {\Omega^I}^*)}{\lambda ({\Omega^I}^* + \omega \Theta^*)} \;, \nonumber \\
	 \kappa &= \frac{(\lambda + 1 + {\Omega^S}^* + \omega)(2+ \omega + {\Omega^I}^*) - \lambda}{\lambda({\Omega^I}^* + \omega \Theta^*)} \;.
\end{align}
This new solution for $\theta_k^*$ expressed in Eq.~\eqref{thetaSol}, despite being more complex than a simple mean-field assumption, provides more flexibility to our approach. 


\paragraph*{Near absorbing-state regime.} At this point, one can already verify the consistency with HMF in the annealed network limit~: Taking $\omega \to \infty$ in Eq.~\eqref{thetaSol}, one recovers $\theta_k^* \to \Theta^*$, illustrating the absence of dynamical correlation. We are left to examine the quasi-static limit, specifically near the absorbing-state ($|\lambda - \lambda_c|/\lambda_c \ll 1$) which dictates our ability to evaluate the threshold. Since $\theta_k^*$ is vanishing $\fa k$ in this regime [see Eq.~\eqref{rhoSS}], we introduce the ratio $\eta_k^* \equiv \theta_k^*/\rho^*$ which rescales the probability $\theta_k^*$ with the prevalence $\rho^* \equiv \sum_k \rho_k^*$.

\begin{figure}
\includegraphics[width = 0.5\textwidth]{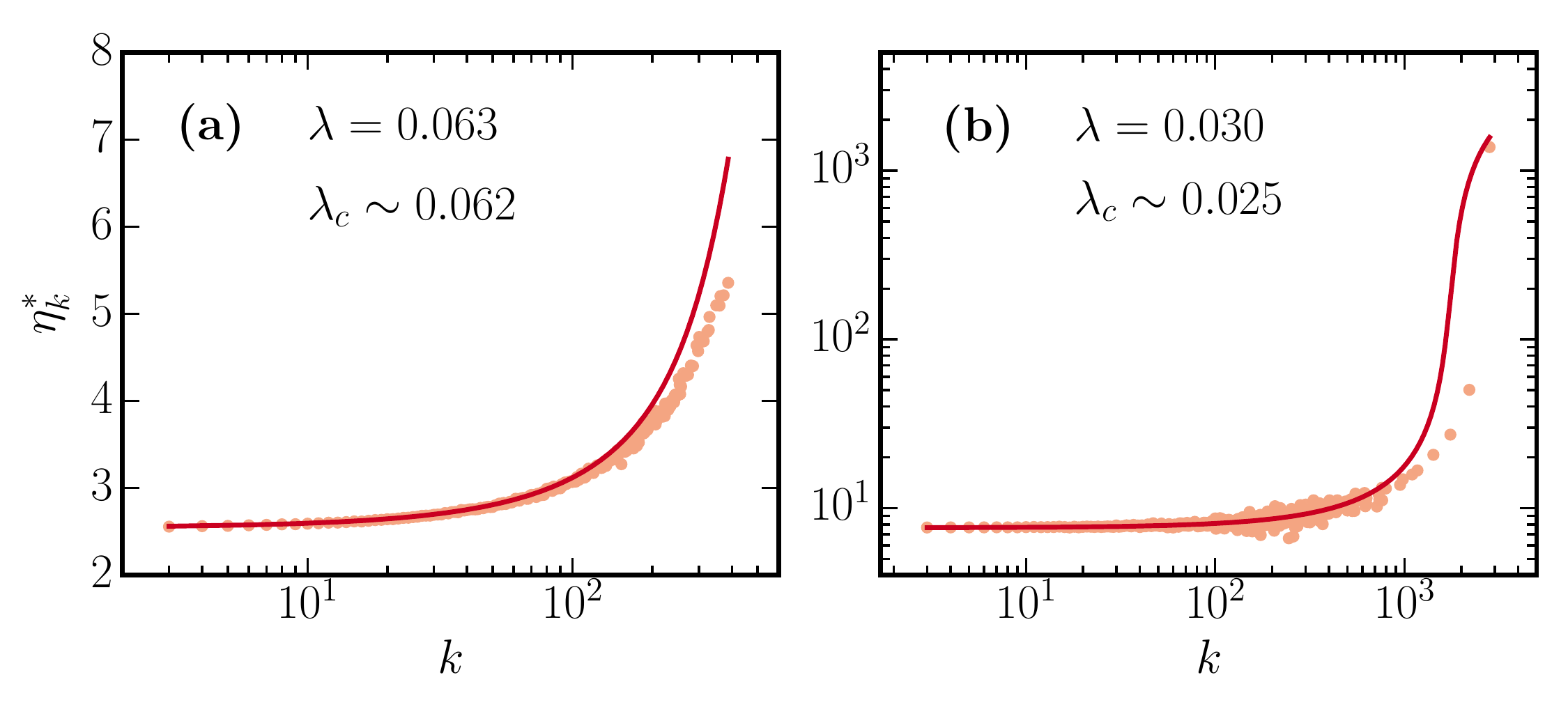}%
\caption{(Color online). Ratio $\eta_k^*$ against the degree $k$ for the SIS dynamics in the quasi-static limit, near the absorbing-state. The solid line corresponds to our estimation from Eq.\eqref{thetaSol} and the markers to Monte-Carlo simulations. Averages are made on multiple rewired realizations of a same degree sequence to simulate the quasi-static limit ($\omega \to 0$). To prevent the system from reaching the absorbing state, we sampled the quasistationary state of the system \cite{Oliveira2005,Ferreira2011,Sander2016}. \textbf{(a)} Collective activation~: $10^2$ realizations with $p_k \sim k^{-\gamma}\exp(-k/k_c)$, $\gamma = 2.5$, $k_c = 100$ and $N = 10^6$. \textbf{(b)} Hub activation~: $5\times 10^3$ realizations with $p_k \sim k^{-\gamma}$, $\gamma = 3.5$ and $N = 10^7$.
 \label{fig:neighborhood}}
\end{figure}

For static networks, it has been observed that there exist two activation schemes depending on the topology~: A collective versus a hub-dominated activation, leading to a delocalized and a localized state in the vicinity of the absorbing-state threshold (see Refs.~\cite{Castellano2012,Mata2015,Ferreira2016}). We observe the same dichotomy in the quasi-static limit---Fig.~\ref{fig:neighborhood} shows that $\eta_k^*$ is qualitatively different in both activation schemes. For the collective one [Fig.~\ref{fig:neighborhood}(a)], $\eta_k^*$ is a growing function of $k$, but it remains of the same order of magnitude over the whole range, while for the hub-dominated one [Fig.~\ref{fig:neighborhood}(b)], $\eta_k^*$ varies over many orders of magnitude. This illustrates the localization of the active state in the neighborhood of the highest degree nodes for the hub-dominated phase transition.

Figure \ref{fig:neighborhood} reveals the importance of using a degree dependent probability $\theta_k^*$ to characterize the quasi-static limit. This also explains the disparity one can observe between the predictions of PHMF and QMF for certain degree distributions---the former approach considers a mean-field probability independent of $k$. Hence, it is unable to describe correctly a hub-dominated activation scheme.

\paragraph*{Absorbing-state threshold.} To take the absorbing-state limit, we start with an active phase ($\lambda > \lambda_c$), then we impose the limit $\lambda \to \lambda_c$, leading to $\rho_k^*, \theta_k^* \to 0 \fa k$. From Eq.~\eqref{thetaSol}, this requires that $\beta \to 0$. It permits us to define the stationary probability $\theta_k^*$ around the critical threshold \cite{SM}
\begin{align} \label{expandedForm}
\theta_k^*(\omega, \lambda \to \lambda_c) &\to \beta f_k(\omega) \;, & f_k(\omega) &\equiv \frac{1}{\widetilde{\kappa}(\omega)-k} \;.
\end{align}
with the parameter
\begin{align}
	\widetilde{\kappa}(\omega) &= \frac{1 + (\lambda_c + 1)^2 + \omega (2 \lambda_c + 3) + \omega^2}{\lambda_c^2} \;.
\end{align}

From Eq.~\eqref{expandedForm},  we impose the constraint $\widetilde{\kappa} > k_{\mathrm{max}}$ to force a positive probability $\theta_k^*$. This constraint leads to the threshold upper bound
\begin{align}\label{upperBound}
	\lambda_c(\omega) < \frac{1 + \omega + \sqrt{2 k_\mathrm{max} -1 + \omega(3 k_\mathrm{max} -1) + \omega^2 k_\mathrm{max}}}{k_{\mathrm{max}}-1} \;.
\end{align}
Equation \eqref{upperBound} points to two major observations. First, with finite rewiring rate $\omega$, our RNA leads explicitly to a vanishing threshold for any random networks in the limit $k_\mathrm{max} \to \infty$. For instance, the quasi-static limit has the particular form
\begin{align}\label{upperBoundQS}
	\lambda_c(\omega \to 0) \equiv \lambda_c^{\mathrm{qs}} < \frac{1 + \sqrt{2k_{\mathrm{max}} -1 }}{k_{\mathrm{max}}-1} \;.
\end{align}
For large $k_\mathrm{max}$, Eq.~\eqref{upperBoundQS} is well approximated by $\lambda_c^{\mathrm{qs}} \lesssim \sqrt{2/k_\mathrm{max}}$, an upper bound previously observed in numerical simulations on static networks and in agreement with pair QMF \cite{Ferreira2012,Mata2013}. Second, it is known that the annealed regime leads to a finite threshold even in the limit $k_\mathrm{max} \to \infty$ for bounded second moment $\avg{k^2}$ \cite{Pastor-Satorras2001}. For this condition to be satisfied, Eq.~\eqref{upperBound} prescribes that the rewiring rate $\omega$ is $\mathcal{O}(\sqrt{k_\mathrm{max}})$. Below this, the highest degree node is able to sustain by itself the dynamics with correlated reinfection from its neighbors \cite{Castellano2010}. In other words, the vanishing of the absorbing-state threshold for any random networks in the limit $k_\mathrm{max} \to \infty$ can be directly attributed to the presence of dynamical correlation, easily tuned in our approach by $\omega$ and in agreement with Ref.~\cite{Boguna2013}.

Combining Eq.~\eqref{expandedForm} with the conservation of the different types of edges (see the Supplemental Material \cite{SM}), we obtain an implicit expression for the absorbing-state threshold, valid for any rewiring rate,
\begin{align}\label{aaThreshold}
	\lambda_c(\omega) &= \frac{(2+\omega)\avg{k f_k}}{(2+\omega)\avg{k^2 f_k}-2\avg{k f_k}} \;.
\end{align}
For arbitrary $\omega$ and degree distribution $p_k$, the quantity $f_k$ is a function of $k$, $\omega$ and $\lambda_c$, thus Eq.~\eqref{aaThreshold} is transcendental and must be solved numerically.

\paragraph*{Limiting cases.} We consider again the extreme regimes of our rewiring process. Equation \eqref{aaThreshold} becomes
\begin{align}\label{aaThresholdLimit}
	\lambda_c &= 
	\begin{dcases}
		\avg{k}/\avg{k^2} & \text{if } \omega \to \infty \;, \\
		\avg{kf_k^{\mathrm{qs}}}/\pr{\avg{k^2f_k^{\mathrm{qs}}}-\avg{kf_k^{\mathrm{qs}}}} & \text{if } \omega \to 0 \;.
	\end{dcases}
\end{align}
where $f_k (\omega \to 0) \equiv f_k^{\mathrm{qs}} $. Hence, we recover as expected the HMF threshold (table \ref{tab:threshold}) in the
annealed limit. In the quasi-static limit, we obtain a threshold similar in form to the one predicted by PHMF (table I), except for the presence of $f_k^{\mathrm{qs}}$ in each average.

\begin{figure}
\includegraphics[width = 0.5\textwidth]{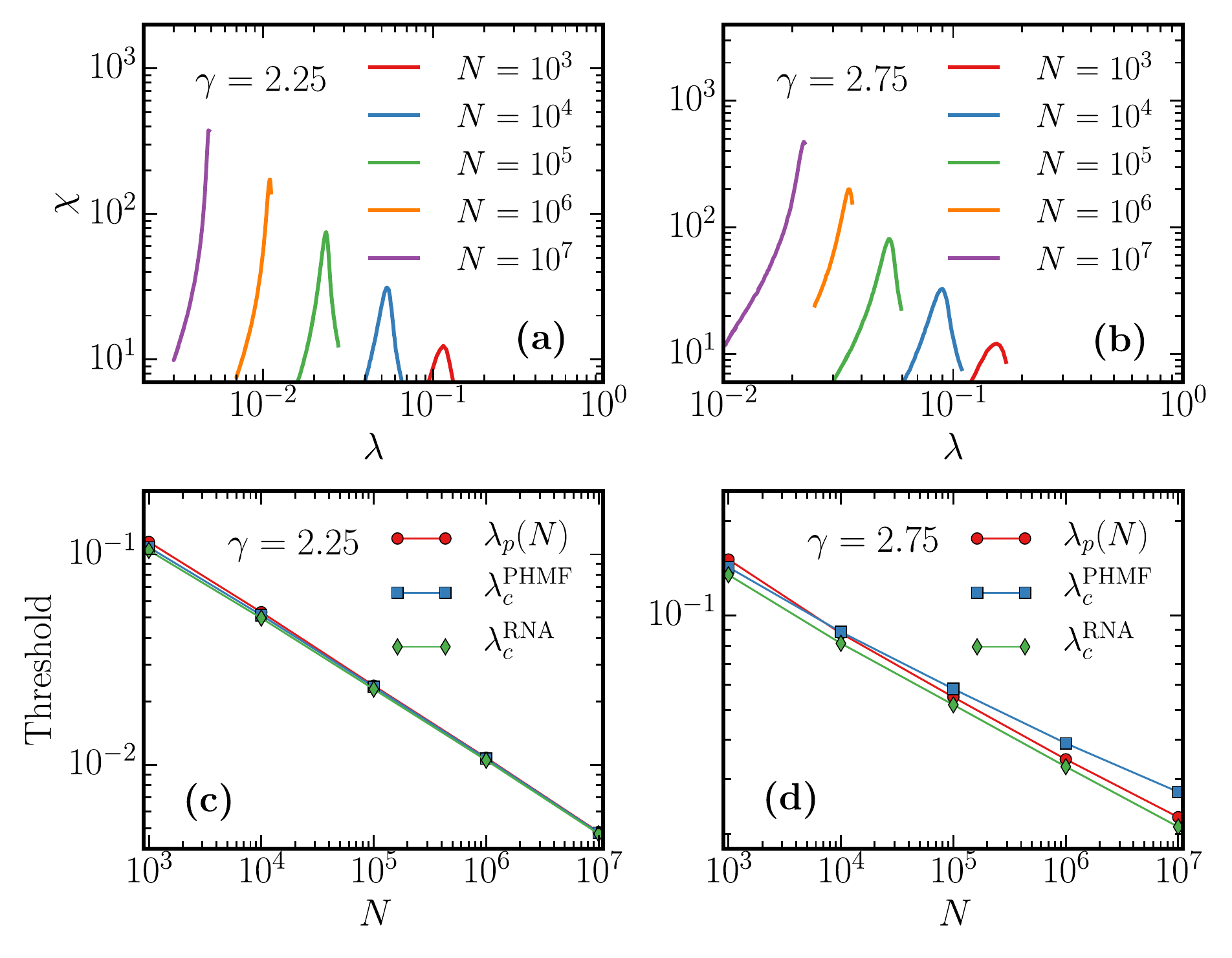}%
\caption{ (Color online). Threshold evaluation for scale-free random networks of degree distribution $p_k \sim k^{-\gamma}$, minimum degree $k_\mathrm{min} = 3$ and maximum degree $k_\mathrm{max}$ bounded by $N^{1/2}$. (a)--(b) Susceptibility against the infection rate for a single network realization. (c)--(d) Threshold against the number of nodes (averaged over 10 network realizations) estimated by~: the position of the susceptibility peak $\lambda_p(N)$, the threshold $\lambda_c^{\mathrm{RNA}}$ of Eq.~\eqref{aaThresholdLimit} for $\omega \to 0$ and the PHMF threshold $\lambda_c^\mathrm{PHMF}$. \label{fig:PL_threshold}}
\end{figure}

It is expected that Eq.~\eqref{aaThresholdLimit} should be a good estimator of $\lambda_c$ for static realizations of the configuration model with large $N$. The validation is presented in Fig.~\ref{fig:PL_threshold}~: in sampling the quasistationary state \cite{Oliveira2005,Ferreira2011,Sander2016}, we evaluate the susceptibility
\begin{align}
\chi &= \frac{\avg{n^2}-\avg{n}^2}{\avg{n}} \;,
\end{align}
with $n \leq N$ the number of infected nodes in the system. The susceptibility exhibits a sharp maximum at $\lambda_p(N)$ as shown in Fig.~\ref{fig:PL_threshold}(a) and \ref{fig:PL_threshold}(b), corresponding to the epidemic threshold of the system in the thermodynamic limit \cite{Ferreira2012}. Figures \ref{fig:PL_threshold}(c) and \ref{fig:PL_threshold}(d) show that our RNA yields an improvement compared to PHMF for scale-free distribution $p_k \sim k^{-\gamma}$ with $\gamma > 5/2$. As pointed out in Refs.~\cite{Castellano2012,Goltsev2012}, $\gamma = 5/2$ separates the collective ($\gamma \leq 5/2$) from the hub-dominated ($\gamma > 5/2$) activation scheme. Our formalism therefore yields the correct threshold for both types of phase transition.

\begin{figure}
\includegraphics[width = 0.40\textwidth]{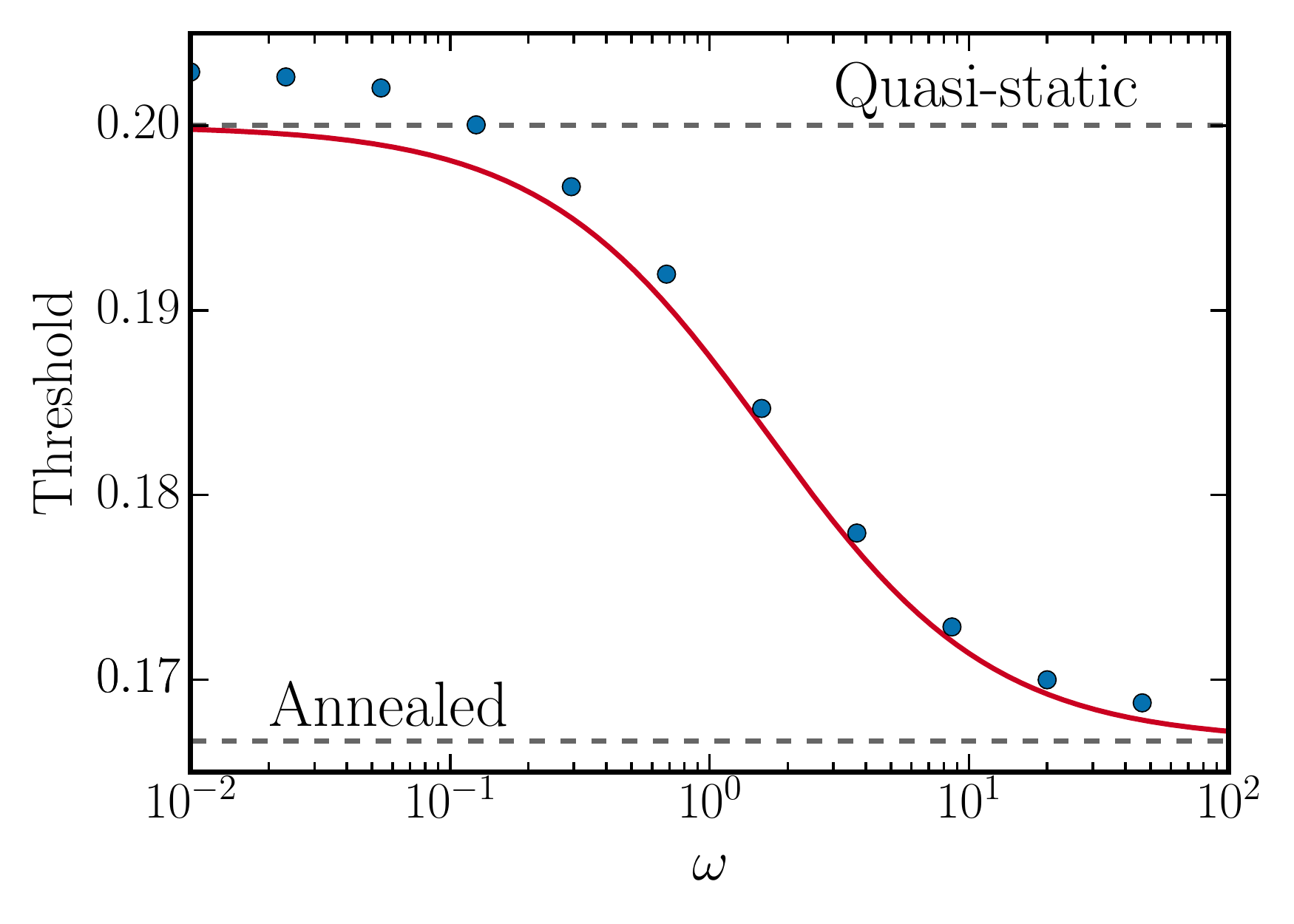}%
\caption{(Color online). Threshold against the rewiring rate for a regular random network with degree $k_0 = 6$ and network size $N = 10^5$. The solid line represents the threshold $\lambda_c^{\mathrm{RNA}}$ of Eq.~\eqref{threshold_RRN} and the markers represent the position of the susceptibility peak $\lambda_p(N)$. The disparity between simulations and the RNA is attributed to finite size effects \cite{Ferreira2012} and to the approximations leading to Eq.~\eqref{aaThreshold}. \label{fig:RRN_threshold}}
\end{figure}

Finally, we are left to examine the regimes between the annealed and the quasi-static limit. To do so, we have extended the standard quasistationary state method to include the rewiring procedure \cite{SM}. We have applied it to a regular random network with distribution $p_k = \delta_{kk_0}$, for which Eq.~\eqref{aaThreshold} yields the threshold
\begin{align}\label{threshold_RRN}
	\lambda_c(\omega) &= \frac{2 + \omega}{(2+\omega)k_0 - 2} \;.
\end{align}
The validation of the RNA is presented in Fig.~\ref{fig:RRN_threshold}, where it is seen that Eq.~\eqref{threshold_RRN} reproduces with very good accuracy the transition from one regime to another.

In summary, our stationary state analysis describes accurately the SIS dynamics on random networks with an arbitrary degree distribution for any rewiring rate. This lead to the unification of some current approaches, being coherent with HMF and pair QMF in the annealed and quasi-static limit, obtaining more from less. Moreover, being able to tune the rewiring rate allows us to highlight the critical role of dynamical correlation, notably responsible for the vanishing of the threshold in the limit $k_\mathrm{max} \to \infty$. Beyond these results, our approach opens the way to the detailed characterization of phase transitions in heterogeneous complex networks featuring a rewiring dynamics.

This work may be pursued by the numerical validation of more study cases and further analytical treatments, for instance a detailed analysis of the critical exponents. Due to the generality and versatility of the RNA, it can easily be applied to other binary-state dynamics and lead to substantial improvements. The stationary state analysis can also be extended to cover other rewiring processes and adaptive networks where the topology co-evolves with the dynamics \cite{Gross2006,Gross2009,Marceau2010}.


\begin{acknowledgments}
The authors acknowledge Calcul Qu\'{e}bec for computing facilities, as well as the financial support of the Natural Sciences and Engineering Research Council of Canada (NSERC) and the Fonds de recherche du Quebec --- Nature et technologies (FRQNT). E. Laurence is
grateful to P. Mathieu for financial support.
\end{acknowledgments}

%


\cleardoublepage



\section*{Supplementary material}

# Susceptible-infected-susceptible dynamics on the rewired configuration model
# Supplemental material

Guillaume St-Onge, Jean-Gabriel Young, Edward Laurence, Charles Murphy, and Louis J. Dubé*
*Département de Physique, de Génie Physique, et d'Optique,*
*Université Laval, Québec (Québec), Canada, G1V 0A6*

(Dated: January 6, 2017)

## CONTENTS

---

* Louis.Dube@phy.ulaval.ca

TABLE I. Synthesis of various quantities definition.

| Quantities | Definition |
|---|---|
| $\lambda$ | Infection rate : rate at which the disease is transmitted from an infected node connected to a susceptible node through an edge. |
| $\mu \equiv 1$ | Recovery rate : rate at which infected nodes become susceptible. |
| $\lambda_c$ | Absorbing-state threshold : critical infection rate below which the epidemics cannot persist in the stationary state ($t \to \infty$). |
| $\omega$ | Rewiring rate : rate at which each stub is disconnected and reconnected at random. |
| $\rho_k(t)$ | Probability that a node of degree $k$ is infected. |
| $\theta_k(t)$ | Probability of reaching an infected node following a random edge starting from a degree $k$ susceptible node. |
| $\phi_k(t)$ | Probability of reaching an infected node following a random edge starting from a degree $k$ infected node. |
| $\Theta(t)$ | Probability that a newly rewired stub reaches an infected node. |
| $s_{kl}(t)$ | Probability that a degree $k$ node is susceptible and has $l \leq k$ infected neighbors. |
| $i_{kl}(t)$ | Probability that a degree $k$ node is infected and has $l \leq k$ infected neighbors. |
| $\Omega^S(t)$ | Mean infection rate for the neighbors of susceptible nodes. |
| $\Omega^I(t)$ | Mean infection rate for the neighbors of infected nodes. |

## I. STATIONARY STATE OF THE DYNAMICS

In this section we obtain an explicit solution for $\theta_k^*$, the stationary state probability of reaching an infected node following a random edge starting from a degree $k$ susceptible node. Considering the SIS dynamics on a rewiring network, we introduce the rate equations adapted from Ref. [1]

$$\begin{aligned}
\dot{s}_{kl} =& i_{kl} - \lambda l s_{kl} + [1 + \omega(1-\Theta)]\left[(l+1)s_{k(l+1)} - l s_{kl}\right] \\
&+ (\Omega^S + \omega\Theta)\left[(k-l+1)s_{k(l-1)} - (k-l)s_{kl}\right],
\end{aligned} \quad (1a)$$

$$\begin{aligned}
\dot{i}_{kl} =& \lambda l s_{kl} - i_{kl} + [1 + \omega(1-\Theta)]\left[(l+1)i_{k(l+1)} - l i_{kl}\right] \\
&+ (\Omega^I + \omega\Theta)\left[(k-l+1)i_{k(l-1)} - (k-l)i_{kl}\right],
\end{aligned} \quad (1b)$$

The mean infection rates $\Omega^S(t), \Omega^I(t)$ and the probability $\Theta(t)$ are evaluated by the following expressions

$$\Omega^S = \lambda \frac{\sum_{k,l}(k-l)l s_{kl} p_k}{\sum_{k,l}(k-l) s_{kl} p_k}, \qquad \Omega^I = \lambda \frac{\sum_{k,l} l^2 s_{kl} p_k}{\sum_{k,l} l s_{kl} p_k}, \qquad \Theta = \frac{\sum_k k \rho_k p_k}{\sum_k k p_k}. \quad (2)$$

We can relate the first moments of $s_{kl}$ and $i_{kl}$ to other quantities defined in table I

$$\sum_l s_{kl} = (1-\rho_k), \qquad \sum_l i_{kl} = \rho_k, \qquad \sum_l l s_{kl} = (1-\rho_k)\theta_k k, \qquad \sum_l l i_{kl} = \rho_k \phi_k k. \quad (3)$$

We retrieve the time evolution of the probabilities $\theta_k$ and $\phi_k$ from Eqs. (1) by applying the chain rule

$$\dot{\theta}_k = \frac{\sum_l \dot{s}_{kl} l}{(1-\rho_k)k} + \frac{\dot{\rho}_k \theta_k}{1-\rho_k}, \quad (4a)$$

$$\dot{\phi}_k = \frac{\sum_l \dot{i}_{kl} l}{\rho_k k} - \frac{\dot{\rho}_k \phi_k}{\rho_k}. \quad (4b)$$

Also, the time evolution of $\rho_k$ has the well known form [2, 3]

$$\dot{\rho}_k = -\rho_k + \lambda(1-\rho_k)\theta_k k \,. \tag{5}$$

We explicit the sum over $l$ in Eqs. (4a) and (4b):

$$\dot{\theta}_k = \frac{1}{k(1-\rho_k)} \sum_l \Big\{ -\lambda l^2 s_{kl} + l i_{kl} + [1+\omega(1-\Theta)] \left[ l(l+1)s_{k(l+1)} - l^2 s_{kl} \right]$$
$$+ (\Omega^S + \omega\Theta) \left[ l(k-l+1)s_{k(l-1)} - l(k-l)s_{kl} \right] \Big\} - \frac{\theta_k}{1-\rho_k} \left[ \rho_k - \lambda k \theta_k (1-\rho_k) \right] \,, \tag{6a}$$

$$\dot{\phi}_k = \frac{1}{k\rho_k} \sum_l \Big\{ \lambda l^2 s_{kl} - l i_{kl} + [1+\omega(1-\Theta)] \left[ l(l+1)i_{k(l+1)} - l^2 i_{kl} \right]$$
$$+ (\Omega^I + \omega\Theta) \left[ l(k-l+1)i_{k(l-1)} - l(k-l)i_{kl} \right] \Big\} + \frac{\phi_k}{\rho_k} \left[ \rho_k - \lambda k \theta_k (1-\rho_k) \right] \,. \tag{6b}$$

With Eq. (3), we have the relations

$$\sum_l \left\{ l(l+1)s_{k(l+1)} - l^2 s_{kl} \right\} = -k(1-\rho_k)\theta_k \,, \tag{7a}$$

$$\sum_l \left\{ l(l+1)i_{k(l+1)} - l^2 i_{kl} \right\} = -k\rho_k \phi_k \,, \tag{7b}$$

$$\sum_l \left\{ l(k-l+1)s_{k(l-1)} - l(k-l)s_{kl} \right\} = k(1-\rho_k)(1-\theta_k) \,, \tag{7c}$$

$$\sum_l \left\{ l(k-l+1)i_{k(l-1)} - l(k-l)i_{kl} \right\} = k\rho_k(1-\phi_k) \,. \tag{7d}$$

Using Eqs. (7) in Eqs (6), we find

$$\dot{\theta}_k = -\frac{\lambda}{k(1-\rho_k)} \sum_l \left\{ l^2 s_{kl} \right\} + r_k \phi_k + (\Omega^S + \omega\Theta)(1-\theta_k) - [1+\omega(1-\Theta)]\theta_k - \theta_k \left( r_k - \lambda k \theta_k \right) \,, \tag{8a}$$

$$\dot{\phi}_k = \frac{\lambda}{k\rho_k} \sum_l \left\{ l^2 s_{kl} \right\} - \phi_k + (\Omega^I + \omega\Theta)(1-\phi_k) - [1+\omega(1-\Theta)]\phi_k + \phi_k \left( 1 - \lambda k \theta_k r_k^{-1} \right) \,, \tag{8b}$$

where $r_k \equiv \rho_k/(1-\rho_k)$. One notes that the Eqs. (8a) and (8b) do not depend on $i_{kl}$. In the stationary state, Eq. (5) leads to

$$\rho_k^* = \frac{\lambda k \theta_k^*}{1+\lambda k \theta_k^*} \quad \text{or} \quad \lambda k \theta_k^* = \frac{\rho_k^*}{1-\rho_k^*} = r_k^* \,. \tag{9}$$

Using this result for the Eqs. (8a) and (8b) in the stationary state, we obtain

$$0 = -\frac{\lambda}{k(1-\rho_k^*)} \sum_l \left\{ l^2 s_{kl}^* \right\} + \lambda k \theta_k^* \phi_k^* + (\Omega^{S*} + \omega\Theta^*)(1-\theta_k^*) - [1+\omega(1-\Theta^*)]\theta_k^* \,, \tag{10a}$$

$$0 = \frac{\lambda}{k(1-\rho_k^*)} \sum_l \left\{ l^2 s_{kl}^* \right\} - \lambda k \theta_k^* \phi_k^* + (\Omega^{I*} + \omega\Theta^*)(1-\phi_k^*)\lambda k \theta_k^* - [1+\omega(1-\Theta^*)]\lambda k \theta_k^* \phi_k^* \,. \tag{10b}$$

Adding Eq. (10a) and Eq. (10b), we express $\phi_k^*$ in terms of $\theta_k^*$

$$\phi_k^* = \frac{(\Omega^{S*} + \omega\Theta^*)(1-\theta_k^*) - [1+\omega(1-\Theta^*)]\theta_k^* + \lambda \theta_k^* k (\Omega^{I*} + \omega\Theta^*)}{\lambda \theta_k^* k (\Omega^{I*} + 1 + \omega)} \,. \tag{11}$$

To obtain a closed solution for $\theta_k^*$, we introduce the pair approximation suggested in Ref. [1]

$$\sum_{l=0}^k l^2 s_{kl}^* \approx (1-\rho_k^*) \left[ k\theta_k^* + k(k-1)\theta_k^{*2} \right] \,, \tag{12}$$

3

Combining Eq. (10a),(11) and (12), we obtain the constraint equation

$$0 = \theta_k^{*2}(k-1)\alpha + \theta_k^*(\kappa - k) - \beta \,, \tag{13}$$

with the parameters defined as

$$\alpha = \frac{1 + \omega + \Omega^{I^*}}{\Omega^{I^*} + \omega\Theta^*} \,, \qquad \beta = \frac{(\Omega^{S^*} + \omega\Theta^*)(2 + \omega + \Omega^{I^*})}{\lambda(\Omega^{I^*} + \omega\Theta^*)} \,, \qquad \kappa = \frac{(\lambda + 1 + \Omega^{S^*} + \omega)(2 + \omega + \Omega^{I^*}) - \lambda}{\lambda(\Omega^{I^*} + \omega\Theta^*)} \,. \tag{14}$$

Equation (13) finally leads to the solution

$$\theta_k^*(\omega, \lambda) = \begin{cases} \dfrac{\beta}{\kappa - 1} & \text{if } k = 1 \,, \\ \dfrac{k - \kappa + \sqrt{(k-\kappa)^2 + 4\alpha\beta(k-1)}}{2\alpha(k-1)} & \text{if } k > 1 \,, \end{cases} \tag{15}$$

where the positive sign is chosen to force a positive value for $\theta_k^*$.

## II. CONSERVATION OF THE EDGES

Equation (15) depends upon the mean infection rates $\Omega^{S^*}, \Omega^{I^*}$ and the probability $\Theta^*$. To fix these quantities, we need to impose constraint equations, which might be for instance self-consistency equations such as Eqs. (2). Another constraint on the dynamics is the conservation of the different types of edges. For the SIS model, the type of an edge may be identified by the state of each endpoint. We express $[AB]$ the density of edges per node with endpoints in state $A$ and $B$, where $A, B \in \{S, I\}$. Since $[SI] = [IS]$, we introduce the conservation of edges

$$[SS] + [II] + 2[SI] = \langle k \rangle \,, \tag{16}$$

with $\langle k \rangle \equiv \sum_k p_k k$ the mean degree of the network. According to our degree based approach, these densities in the stationary state are

$$[SS]^* = \sum_k p_k (1 - \rho_k^*)(1 - \theta_k^*)k \,, \tag{17a}$$

$$[II]^* = \sum_k p_k \rho_k^* \phi_k^* k \,, \tag{17b}$$

$$[SI]^* = \sum_k p_k (1 - \rho_k^*)\theta_k^* k \,. \tag{17c}$$

Using Eqs. (9) and (17), the conservation of edges is written

$$\sum_k \left(\frac{p_k}{1 + \lambda\theta_k^* k}\right) \left[\lambda\theta_k^* \phi_k^* k^2 + (1 - \theta_k^*)k + 2\theta_k^* k - (1 + \lambda\theta_k^* k)k\right] = 0 \,,$$

$$\sum_k \left(\frac{p_k \theta_k^* k}{1 + \lambda\theta_k^* k}\right) \left[\lambda\phi_k^* k + 1 - \lambda k\right] = 0 \,. \tag{18}$$

Combining Eq. (18) with Eq. (11), we obtain

$$\sum_k \left(\frac{kp_k}{1 + \lambda\theta_k^* k}\right) \left(\Omega^{S^*} + \omega\Theta^* + \theta_k^* \left\{\Omega^{I^*} - \Omega^{S^*} - \lambda k[1 + \omega(1 - \Theta^*)]\right\}\right) = 0 \,. \tag{19}$$

## III. ABSORBING-STATE LIMIT

To take the absorbing-state limit, we start with an active phase ($\lambda > \lambda_c$) and then impose the limit $\lambda \to \lambda_c$. For coherence in the mathematical development, we assume degree distributions that are bounded by $k_{\max} < \infty$, which

also implies $\lambda_c > 0$. General results for unbounded degree distributions can be obtained afterwards by taking the limit $k_{\max} \to \infty$. Following this assumption, the absorbing-state limit implies $\rho_k^* \to 0 \; \forall k$, or more simply $\theta_k^* \to 0 \; \forall k$. This allows us to identify the behavior of a number of quantities when $\lambda \to \lambda_c$. With the pair approximation Eq. (12), the mean infection rates are rewritten as

$$\Omega^{S^*} = \lambda \frac{\sum_k (1-\rho_k^*)(\theta_k^* - \theta_k^{*2})(k-1)k p_k}{\sum_k (1-\rho_k^*)(1-\theta_k^*) k p_k} \;, \qquad \Omega^{I^*} = \lambda \frac{\sum_k (1-\rho_k^*)[\theta_k^* k + \theta_k^{*2} k(k-1)] p_k}{\sum_k (1-\rho_k^*)\theta_k^* k p_k} \;. \tag{20}$$

Therefore, in the limit $\theta_k^* \to 0 \; \forall k$, we have

$$\lim_{\lambda \to \lambda_c} \Omega^{S^*} = 0 \;, \qquad \lim_{\lambda \to \lambda_c} \Omega^{I^*} = \lambda_c \;, \qquad \lim_{\lambda \to \lambda_c} \Theta^* = 0 \;. \tag{21}$$

From Eq. (15), the limit $\theta_k^* \to 0 \; \forall k$ implies the following constraint

$$\lim_{\lambda \to \lambda_c} \kappa > k \; \forall k \;. \tag{22}$$

The limit $\Omega^{S^*}, \Theta^* \to 0$ also implies $\beta \to 0$ from its definition [Eq. (14)]. To compare the behavior of each observable as $\lambda \to \lambda_c$, we develop them in terms of $\beta \ll 1$. First, we expand Eq. (15) for $k > 1$

$$\theta_k^*(\omega, \lambda \sim \lambda_c) = \frac{k - \kappa + |k-\kappa| + \frac{2\alpha\beta(k-1)}{|k-\kappa|}}{2\alpha(k-1)} + \mathcal{O}(\beta^2)$$

$$= \frac{\beta}{\kappa - k} + \mathcal{O}(\beta^2) \tag{23}$$

where the second line follows from Eq. (22). We can also rewrite $\kappa$ as

$$\kappa = \underbrace{\frac{\lambda_c^2 + (\omega+1)[2\lambda_c + (\omega+2)]}{\lambda_c^2}}_{\widetilde{\kappa}} + \mathcal{O}(\beta) \tag{24}$$

which leads to the expression

$$\theta_k^*(\omega, \lambda \sim \lambda_c) = \beta f_k(\omega) + \mathcal{O}(\beta^2) \tag{25}$$

where $f_k(\omega) \equiv (\widetilde{\kappa} - k)^{-1}$. One notes that this expression also hold for $k=1$ [see Eq. (15)]. With Eq. (25), we have the following leading terms near the absorbing-state

$$\Omega^{S^*} = \mathcal{O}(\beta) \;, \qquad \Omega^{I^*} = \lambda_c + \mathcal{O}(\beta) \;, \qquad \Theta^* = \mathcal{O}(\beta) \;. \tag{26}$$

We can also relate $\Omega^{S^*}$ and $\Theta^*$ using the definition of $\beta$ [Eq. (14)]

$$\Omega^{S^*} = \frac{\beta \lambda_c^2}{\lambda_c + 2 + \omega} - \omega \Theta^* + \mathcal{O}(\beta^2) \tag{27}$$

To obtain a closed form expression of $\lambda_c$, we combine Eq. (27) with the conservation of the edges Eq. (19) to obtain

$$\beta \sum_k p_k k \left\{ \frac{\lambda_c^2}{\lambda_c + 2 + \omega} + f_k \lambda_c [1 - k(1+\omega)] \right\} + \mathcal{O}(\beta^2) = 0 \;. \tag{28}$$

This equation must hold for any value of $\beta$ as $\beta \to 0$, hence we obtain the following constraint on $\lambda_c$

$$\sum_k p_k k f_k \left\{ \frac{(\widetilde{\kappa} - k)\lambda_c^2}{\lambda_c + 2 + \omega} + \lambda_c [1 - k(1+\omega)] \right\} = 0 \;. \tag{29}$$

Rearranging the terms in curly braces and factorizing, Eq. (29) becomes

$$\sum_k p_k k f_k \left\{ 2 + \omega + \lambda_c [2 - (2+\omega)k] \right\} = 0 \;, \tag{30}$$

and is finally rewritten under the form

$$\lambda_c = \frac{(2+\omega)\langle k f_k \rangle}{(2+\omega)\langle k^2 f_k \rangle - 2 \langle k f_k \rangle} \;, \tag{31}$$

where averages are taken over $p_k$.

## IV. NUMERICAL SIMULATIONS

To simulate the SIS dynamics on networks, we used a modified Gillespie algorithm [4]. During the simulation process, we track the number of infected nodes $N_I(t)$ and the number of stubs emanating from them $Q_I(t)$. The total number of stubs is denoted $Q$ and is fixed according to our rewiring process. At each step, three event types are possible with the following probability

$$p = \begin{cases} N_I(t)/[N_I(t) + \lambda Q_I(t) + \omega Q/4] & \text{Recovery event}, \\ \lambda Q_I(t)/[N_I(t) + \lambda Q_I(t) + \omega Q/4] & \text{Infection attempt}, \\ [\omega Q/4]/[N_I(t) + \lambda Q_I(t) + \omega Q/4] & \text{Rewiring event}. \end{cases} \quad (32)$$

Each event occurs as follows

- Recovery event : an infected node is chosen randomly and becomes susceptible.
- Infection attempt : an infected node is chosen proportionally to its degree, then we choose one of its emanating stubs randomly for an infection. If the node at the other endpoint is infected, we do nothing : this frustrated event corrects the probability in order to make the process equivalent to randomly choosing an edge among the set of all $S-I$ edges.
- Rewiring event : Two edges $(a_1, b_1)$ and $(a_2, b_2)$ are randomly chosen with $a_i, b_i$ the labels for the nodes; choosing an edge $(b_1, a_1)$ is equally likely. We then rematch the stubs according to the following scheme $(a_1, b_1), (a_2, b_2) \mapsto (a_1, b_2), (a_2, b_1)$. Loops and multi-edges are permitted.

After all events—even the frustrated ones—we update the time with $t \mapsto t + \Delta t$ where $\Delta t \equiv \langle \Delta t \rangle = [N_I(t) + \lambda Q_I(t) + \omega Q/4]^{-1}$. Choosing the average value $\langle \Delta t \rangle$ instead of drawing $\Delta t$ from an exponential distribution here does not affect our results due to the large averaging time used.

In order to evaluate some dynamic observables for infection rate $\lambda$ near the absorbing-state threshold, we sample the configurations of the system that do *not* fall on the absorbing-state (every node is susceptible)—the quasi-stationary state (QS) [5–7]. Each time the system tries to visit the absorbing-state, the actual state is replaced by a configuration randomly chosen among a list of $M$ previously stored active configurations. Also, with probability $\alpha \Delta t$, each active configuration is stored, thus updating the list proportionally to their average lifetime. The system is then expected to converge on the QS state over which we measure observables. In all our simulations, we chose $M \in [50, 100]$ and a storing factor of $\alpha = 10^{-2}$. For $\omega > 0$, we kept in memory the network structure and a vector describing the state of each node.

---



\end{document}